\documentclass[12pt]{iopart}

\usepackage{amsfonts}
\input epsf

\def\bce{\begin{center}}
\def\ece{\end{center}}
\def\beq{\begin{eqnarray}}
\def\eeq{\end{eqnarray}}
\def\ben{\begin{enumerate}}
\def\een{\end{enumerate}}
\def\bei{\begin{itemize}}
\def\eei{\end{itemize}}

\def\nn{\nonumber}

\def\tr{\mbox{tr}}

\def\brr{\begin{array}}
\def\err{\end{array}}

\begin{document}

\title[]{Zeta Function Methods and Quantum Fluctuations\footnote{Talk given at the
Conference ``Quantum Theory and Symmetries - 5'', Valladolid
(Spain), July 22 - 28, 2007}}

\author{Emilio Elizalde}

\address{Instituto de Ciencias del Espacio (CSIC) \\
Institut d'Estudis Espacials de Catalunya (IEEC/CSIC) \\
Campus UAB, Facultat de Ci\`encies, Torre C5-Parell-2a planta \\
E-08193 Bellaterra (Barcelona) Spain} \ead{elizalde@ieec.uab.es
http://www.ieec.fcr.es/english/recerca/ftc/eli/eli.htm}

\begin{abstract}
A review of some recent advances in zeta function techniques is
given, in problems of pure mathematical nature but also as applied
to the computation of quantum vacuum fluctuations in different field
theories, and specially with a view to cosmological applications.

\end{abstract}

\maketitle

\section{Introduction}
Zeta function regularization methods are optimally suited for the
calculation of the contribution of fluctuations of the vacuum
energy, of the quantum fields pervading the universe, to the
cosmological constant. Order of magnitude calculations of the
absolute contributions of all fields are known to lead to a value
which is off by over hundred and twenty orders, as compared with the
results obtained from observational fits, what is known as the {\it
new cc problem}. This is difficult to solve and many authors still
stick to the old problem to try to prove that basically its value is
zero with some perturbations thereof leading to the (small) observed
result (Burgess et al., Padmanabhan, etc.) We have also addressed
this issue recently in a somewhat similar way, by considering the
{\it additional} contributions to the cosmological constant that may
come from the possibly non-trivial topology of space and from
specific boundary conditions imposed on braneworld and other
seemingly reasonable models that are being considered in the
literature (mainly with other purposes too). This kind of Casimir
effect would play at a cosmological scale. If the ground value of
the cc would be indeed zero (and there are different hints pointing
out towards this), we could then be left with this perturbative
quantity coming from the topology or boundary conditions and, in
particular it could be the fact that the computed number is of the
right order of magnitude (and has the right sign, what is also
non-trivial) when compared with the observational value. This is
proven to be true in some of the aforementioned examples. A further
step in this approach would be to consider the so-called dynamical
Casimir effect or Davies-Fulling theory. Although there is no clear
understanding of how it should be applied in cosmology, some
considerations regarding its correct renormalization at laboratory
scales have been made recently and we will refer to them later.

The ones above are the physical issues we would like to address
ultimately. This needs first the heavy mathematics of zeta
functions. They will be presented in the first part of this work in
fair detail. The paper is organized as follows, in correspondence
with the material presented at the Conference. As a tribute to the
actual discoverer of the zeta function, namely Leonhard Euler, in
this Celebration Year, Sect. 2 recalls some essential points that
lead him to introduce this function---widely considered to be the
most important function in Mathematics---with a quick view over the
many extensions of that concept in the following centuries. In Sect.
3 we describe how the concept of zeta function of a
pseudodifferential operator has become a decisive tool for the
regularization of quantum field theories, in special in curved
space-time, as clearly realized by S. Hawking. This is exemplified
in Sect. 4 through the regularization of the vacuum fluctuations of
a quantum system, under some boundary conditions, with a reference
to the case of the dynamical Casimir effect (moving boundaries),
where regularization issues are particularly involved. Finally,
Sect. 5 is devoted to the possible applications of these results in
cosmology, concerning the dark energy issue.

\section{Euler and the Zeta Function}
There are beautiful accounts on how Euler discovered the zeta
function (see, e.g. \cite{ayo1,watkin}). The  harmonic series
\begin{eqnarray}  H = 1 + \frac{1}{ 2} + \frac{1}{3} + \frac{1}{4} +
\frac{1}{5} + \frac{1}{6} + \cdots \end{eqnarray} was well known to
have an infinite sum. Euler asked himself about the `prime harmonic
series'
\begin{eqnarray} PH = 1+ \frac{1}{2} + \frac{1}{ 3} + \frac{1}{5} + \frac{1}{7} +
\frac{1}{11} +\cdots, \end{eqnarray}  is it finite or infinite? It
is a fact that one cannot split the first series into two, one of
them being the second, as
\begin{eqnarray}  \left( 1 + \frac{1}{ 2} + \frac{1}{3} + \frac{1}{5} +
\frac{1}{7} +\cdots \right)+ \left( \frac{1}{ 4} + \frac{1}{6} +
\frac{1}{8} + \frac{1}{9} +\frac{1}{10} +\cdots
\right)\end{eqnarray} and try to show that the second is finite
(what would mean  the first part is infinite). So Euler considered
the function
\begin{eqnarray}  \zeta (s) = 1+ \frac{1}{2^s} + \frac{1}{3^s} +
\frac{1}{4^s} + \frac{1}{5^s} +\cdots\end{eqnarray} Provided $s$ is
bigger than $1$, one can certainly split it up as \begin{eqnarray}
\hspace*{-12mm} \left( 1 + \frac{1}{ 2^s} + \frac{1}{3^s} +
\frac{1}{5^s} + \frac{1}{7^s} +\cdots \right) + \left( \frac{1}{
4^s} + \frac{1}{6^s} + \frac{1}{8^s} + \frac{1}{9^s} +\frac{1}{10^s}
+\cdots \right). \end{eqnarray} Now the idea is to prove that when
$s$ approaches $1$ the first sum becomes divergent. Thus this power
$s$ was very useful.

Making things short, a key step in the whole argument is the
celebrated factorization of the whole zeta function in terms of
prime numbers, namely \begin{eqnarray} \hspace*{-12mm} \zeta (s) =
\frac{1}{ 1 - 1/2^s} \times \frac{1}{ 1 - 1/3^s } \times\frac{1}{ 1
- 1/5^s} \times\frac{1}{ 1 - 1/7^s} \times\frac{1}{ 1 - 1/11^s}
\times \cdots
\end{eqnarray} This comes from the fact that for any prime  $p$  and
any power $s > 1$, setting $x = 1/p^s$ one has the geometric series
\begin{eqnarray} \frac{1}{ 1 - 1/p^s} = 1+ \frac{1}{ p^s} +
\frac{1}{ p^{2s}} + \frac{1}{ p^{3s}} + \cdots \end{eqnarray} Euler
multiplied together these infinite sums to express his infinite
product as a single infinite sum as \begin{eqnarray}  \frac{1}{
p_1^{k_1s} \cdots p_n^{k_ns}}, \end{eqnarray} with $p_1, \ldots ,
p_n$  primes, $ k_1,\ldots, k_n $ positive integers, each such
combination occurs exactly once and the rhs is just a rearrangement
of $\zeta (s)$. It is widely recognized nowadays that Euler's {\it
infinite product formula} for $\zeta (s)$ marked the beginning of
{\it analytic number theory.}

Dirichlet modified the {\it zeta function} introduced by Euler.
Primes were separated into categories, depending on the {\it
remainder} when divided by $k$:  \begin{eqnarray} L(s, \chi ) =
\frac{\chi (1)}{1^s} + \frac{\chi (2)}{2^s} + \frac{\chi (3)}{3^s} +
\frac{\chi (4)}{4^s} + \cdots, \end{eqnarray} where $\chi (n)$ is a
special function now known as a Dirichlet {\it `character'}, that
splits the primes in the required way. It satisfies the conditions:
\ben \item   $ \chi (mn) = \chi (m)\chi (n),$ for any $  m, n;$
\item $\chi (n)= \chi  (n + k), \ \forall  n;$  \item   $\chi (n) =
0$, if $n, k$ have a common factor; \item   $\chi (1) = 1$. \een Any
function $L(s, \chi ),$ where $s$ is a real number bigger than $1$
and $\chi $ a character, is known as a Dirichlet $L$-series. The
Euler zeta function is the special case with  $\chi (n) = 1$ for all
$n$, another example being $\chi (n) = \mu (n)$ (the M\"{o}bius
function).

A very crucial generalization, introduced by Bernhard Riemann, was
to allow  $s$ and  $\chi (n)$ to be {\it complex.} The celebrated
{\it Riemann zeta function,} subsequently extended by Hurwitz,
Lerch, Epstein, Barnes, etc. increased the number and importance of
the zeta function concept decisively. Many results about prime
numbers were proven and $L$-series provide still now a powerful tool
for the study of the primes. We should mention for completeness that
the concept of zeta function has been yet much more extended, first
to the concept of zeta function of a pseudifferential operator (as
we are going to see next), but also to the orbits and trajectories
in dynamical systems, under the form of the Selberg zeta function,
the Ruelle, the Lefschetz zeta function, and many others that lie
outside the scope of this brief summary (Arakelov geometry is one of
the most active developments right now). In Ref. \cite{watkin} a
directory of all known zeta functions can be found (there is even
one named after the author of the present article, see also Keith
Devlin's account there).

\section{The Zeta Function of a Pseudodifferential Operator}
A {\it pseudodifferential operator} $A$ of order $m$ on a manifold
$M_n$ is defined through its symbol $a(x,\xi)$, which is a  function
belonging to the  space $S^m(\mathbb{R}^n\times \mathbb{R}^n)$ of
$\mathbb{C}^\infty$ functions such that for any pair of multi-indexs
$\alpha, \beta$ there exists a constant $C_{\alpha,\beta}$ so that $
\left| \partial^\alpha_\xi
\partial^\beta_x a(x,\xi) \right| \leq
 C_{\alpha,\beta} (1+|\xi|)^{m-|\alpha|}.
$ The definition of $A$ is given, in the distribution sense, by
\begin{equation}
Af(x) = (2\pi)^{-n} \int e^{i<x,\xi>} a(x,\xi) \hat{f}(\xi) \, d\xi,
\end{equation}
$f$ a smooth function, $f \in
 {\cal S}$, recall  $
{\cal S} = \left\{ f \in  C^\infty (\mathbb{R}^n);  \mbox{sup}_x
|x^\beta \partial^\alpha f(x) | < \infty, \forall \alpha, \beta \in
\mathbb{R}^n\right\}$,
 $ {\cal S}'$ being the space of tempered
distributions and $\hat{f}$ the Fourier transform of $f$.
 When $a(x,\xi)$ is a polynomial in $\xi$
 one gets a differential operator.
In general,  the order $m$ can be complex. The {\it symbol} of a
$\Psi$DO  has the form $ a (x,\xi) =a _m(x,\xi) +a _{m-1}(x,\xi) +
\cdots +a _{m-j}(x,\xi) + \cdots, \label{spsd} $ being $a _k(x,\xi)
= b_k(x) \, \xi^k$. The symbol $a(x,\xi)$ is said to be {\it
elliptic} if it is invertible for large $|\xi |$ and if there exists
a constant $C$ such that $|a(x,\xi)^{-1}| \leq C (1+ |\xi |)^{-m}$,
for $|\xi | \geq C$. An  elliptic $\Psi$DO is one with an elliptic
symbol.

Pseudodifferential operators [$\Psi$DO] are useful tools, both in
mathematics and in physics. They were crucial for the proof of the
uniqueness of the Cauchy problem \cite{cald} and also for the proof
of the Atiyah-Singer index formula \cite{as63}. In quantum field
theory they appear in any analytical continuation process (as
complex powers of differential operators, like the Laplacian)
\cite{seel1}. And they constitute nowadays the basic starting point
of any rigorous formulation of quantum field theory \cite{fw2}
through microlocalization, a concept that is considered to be the
most important step towards the understanding of linear partial
differential equations since the invention of distributions
\cite{psdo}.

\subsection{Definition of the Zeta Function}
Let $A$ a positive-definite elliptic $\Psi$DO of positive order $m
\in \mathbb{R}$, acting on the space of smooth sections of $E$, an
$n$-dimensional vector bundle over $M$, a  closed $n$-dimensional
manifold. The {\it zeta function} $\zeta_A$ is defined as \beq
\zeta_A (s) = \mbox{tr}\ A^{-s} = \sum_j
 \lambda_j^{-s}, \qquad \mbox{Re}\ s>\frac{n}{m} \equiv s_0,
\eeq where $s_0=$ dim$\,M/$ord$\,A$ is called the {\it abscissa of
convergence} of $\zeta_A(s)$.  Under these conditions, it can be
proven that $\zeta_A(s)$ has a meromorphic continuation to  the
whole complex plane $\mathbb{C}$ (regular at $s=0$), provided that
the principal symbol of $A$ (that is $a_m(x,\xi)$) admits a {\it
spectral cut}: $
 L_\theta = \left\{  \lambda \in \mathbb{C};
 \mbox{Arg}\, \lambda =\theta,
\theta_1 < \theta < \theta_2\right\},   \mbox{Spec}\, A \cap
L_\theta = \emptyset $ (Agmon-Nirenberg condition).
 The definition of $\zeta_A (s)$ depends on the
position of the cut $L_\theta$.
 The only possible singularities of $\zeta_A (s)$ are
{\it poles} at $ s_k = (n-k)/m,  \  k=0,1,2,\ldots,n-1,n+1, \dots. $
 M. Kontsevich and S. Vishik have managed to extend this
definition to the case when  $m \in \mathbb{C}$ (no spectral
 cut exists) \cite{kont95b}.

\subsection{$\Psi$DOs on Boundaryless Manifolds}
Let $M$ be a compact $n$-dim $C^\infty$ manifold without a boundary,
$E$ a smooth Hermitian vector bundle over $M$, $A$ a positive
$\Psi$DO of positive order $m$ in $E$, elliptic and selfadjoint
(admissible). The operator $e^{-tA}$, namely  $e^{-tA}: f \mapsto
u$, is the solution operator for the heat equation: $\partial_t u +A
u=0$, with initial value $  \left. u\right|_{t=0} =f$.

 This operator is traceclass  $  \forall t> 0$,  and as \ $  t\downarrow
0$ it satisfies \beq \tr e^{-tA} \sim \sum_{j=0}^\infty \alpha_j (A)
t^{(j-n)/m} + \sum_{k=1}^\infty \beta_k (A) t^k \log t. \eeq By
Mellin transform: \beq  \zeta_A (s) = \frac{1}{\Gamma (s) }
\int_0^\infty e^{-tA}\, t^{s-1}\, dt,\eeq
 $  \zeta_A (s)$ has a meromorphic extension with only possible poles
at  $  s_j =(n-j)/m$, $  j\in \mathbb{N}$,  at most {\it simple} at
$ s_j \notin -\mathbb{N}$, and at most {\it  double} at  $  s_j \in
-\mathbb{N}$. Moreover,  \beq \alpha_j (A)  = \mbox{Res}_{s=s_j}
\Gamma (s) \zeta_A (s), \qquad \beta_k (A)  = \mbox{Res}_{s=-k}
(s+k) \Gamma (s) \zeta_A (s)\eeq
 The {\it  asymptotic expansion} of the heat kernel determines the
{\it  pole structure} of $  \zeta_A (s)$,  and vice versa. (i) If $
A$ is a differential operator, then:   $  \alpha_j (A) =0,\, j$ odd,
$ \beta_k (A) =0, \forall k$. (ii) If $  A \geq 0$ one still has the
same results, but now  for $
A-$Ker$A$ (subtract DimKer to the residue at $  0$). (iii) If $
s_j\in \mathbb{N}$, then $  \alpha_j (A) $ is not locally computable
\cite{gg1,cvz1}.

\subsection{The Zeta Determinant}
Let $A$ a $\Psi$DO  operator with a  spectral decomposition: $ \{
\varphi_i, \lambda_i \}_{i\in I}$, where $I$ is some set of indices.
The definition of determinant \cite{soulevoros} starts by trying to
make sense of the product   $  \prod_{i\in I} \lambda_i$, which can
easily be transformed into a `sum':
   $ \ln \prod_{i\in I} \lambda_i   = \
      \sum_{i\in I} \ln \lambda_i $. From  the
definition of the  zeta function of $A$:  $\zeta_A (s) =
    \sum_{i\in I} \lambda_i^{-s}$, by
 taking the derivative at $s=0$:   $\zeta_A ' (0) =  - \sum_{i\in I}
   \ln \lambda_i $, we arrive at the following definition of
determinant of $A$ \cite{rs1}:
  \begin{equation} {\det}_\zeta A = \exp \left[ -\zeta_A ' (0)
   \right] .\end{equation}
An older definition (due to Weierstrass) is obtained by subtracting
in the series above (when it is such) the leading behavior of
 $\lambda_i$ as a function of  $i$, as $i \rightarrow \infty$,
 until the series
 $ \sum_{i\in I} \ln \lambda_i $ is made to converge. The shortcoming
 is here---for physical applications---that these additional terms turn out to be
{\it non-local} in general and, thus, they are non-admissible in a
renormalization procedure \cite{renorm1}.

In algebraic QFT, in order to write down an action in operator
language one needs a functional that replaces integration.
 For the Yang-Mills theory this is the Dixmier trace, which
 is the {\it unique} extension of the usual trace to the ideal
 ${\cal L}^{(1,\infty)} $
of the compact operators  $T $ such that the partial sums of its
spectrum diverge logarithmically as the number of terms in the sum:
$ \sigma_N (T) \equiv \sum_{j=0}^{N-1} \mu_j= {\cal O} (\log N), \
\mu_0 \geq \mu_1 \geq \cdots$
 The definition of the Dixmier trace of $T$ is: Dtr $T =
 \lim_{N\rightarrow \infty}
\frac{1}{\log N} \sigma_N (T),$ provided that the Cesaro means
$M(\sigma) (N)$
 of the sequence in $N$ are convergent as $N
 \rightarrow \infty$
[remember that: $M(f)(\lambda) =\frac{1}{\ln \lambda} \int_1^\lambda
f(u) \frac{du}{u}$].
 Then, the Hardy-Littlewood theorem can be stated in a way
 that connects the Dixmier trace with the residue of the zeta
function of the operator $T^{-1}$ at $s=1$ (see Connes
\cite{conn1}): $ \mbox{Dtr}\ T= \lim_{ s \rightarrow  1^+} (s-1)
\zeta_{T^{-1}} (s). $

\subsection{The Wodzicki Residue}
The Wodzicki (or noncommutative) residue
 \cite{wodz87b}
 is the {\it only} extension of the Dixmier trace to the
$\Psi$DOs which are not in  ${\cal L}^{(1,\infty)}$. It is the {\it
only}
 trace  one can define in the algebra of $\Psi$DOs (up to a
multiplicative constant),
 its definition being: res $A=2$ Res$_{s=0}\ \tr (A \Delta^{-s})$,
with  $\Delta$ the Laplacian. It satisfies the trace condition: res
$(AB)=$ res $(BA)$. A very important property is that  it can be
expressed as an integral (local form) $ \mbox{res} \ A = \int_{S^*M}
\mbox{tr}\   a_{-n}(x,\xi)
 \, d\xi
$ with $S^*M \subset T^*M $ the co-sphere bundle on $M$ (some
authors put a coefficient  in front of the integral: Adler-Manin
residue).

If dim $M=n=-$  ord $A$ ($M$ compact Riemann, $A$ elliptic, $n\in
\mathbb{N}$) it coincides with
 the Dixmier trace, and one has
$ \mbox{Res}_{s=1} \zeta_A (s) = \frac{1}{n} \, \mbox{res} \ A^{-1}.
$
 The Wodzicki residue continues to make sense
for $\Psi$DOs of arbitrary order and, even if the symbols $a_{j} (x,
\xi)$, $j<m$, are not invariant under coordinate choice, their
integral is, and defines a trace. All residua at poles of the zeta
function of a $\Psi$DO can be easily obtained from the Wodzciki
residue \cite{6a}.

\subsection{Singularities of $\zeta_A$}
A complete determination of the meromorphic
 structure of some zeta  functions in the complex plane can be
also obtained by means of the Dixmier trace and the Wodzicki
residue. Missing for the full description of the singularities in
the above are just the {\it  residua} of all the poles. As for the
regular part of the analytic continuation, specific methods have to
be used (see later). It can be proven that, under the conditions of
existence of the zeta function of $  A$, given above, and being the
 symbol  $  a(x,\xi)$ of the operator $  A$  analytic
in $  \xi^{-1}$ at  $  \xi^{-1}=0$, then it follows that \beq
\mbox{Res}_{s=s_k} \zeta_A (s) = \frac{1}{m} \, \mbox{res} \
A^{-s_k} =  \frac{1}{m} \int_{S^*M} \mbox{tr}\ a^{-s_k}_{-n} (x,\xi)
\, d^{n-1} \xi. \eeq  The proof is rather simple and it can be
obtained by invoking the homogeneous component of degree  $ -n$ of
the corresponding power of the principal symbol of  $ A$, obtained
by the appropriate derivative of  a power of the symbol with respect
to $  \xi^{-1}$ at  $  \xi^{-1}=0$, namely \beq  a^{-s_k}_{-n}
(x,\xi) = \left.  \left( \frac{\partial}{\partial
 \xi^{-1}} \right)^k \left[ \xi^{n-k} a^{(k-n)/m} (x,\xi)
\right] \right|_{ \xi^{-1} =0}  \xi^{-n}. \eeq Then the proof
follows constructively, by easy algebraic manipulation.

\subsection{The Multiplicative Anomaly and its Implications}
Given $A$, $B$ and $AB$ $\Psi$DOs, even if $\zeta_A$, $\zeta_B$ and
$\zeta_{AB}$ exist, it turns out that, in general, $ {\det}_\zeta
(AB) \neq {\det}_\zeta A \ {\det}_\zeta B. $
 The multiplicative (or noncommutative) anomaly (or defect) is defined as:
\beq \delta (A,B) = \ln \left[ \frac{ \det_\zeta (AB)}{\det_\zeta A
\ \det_\zeta B} \right] = -\zeta_{AB}'(0)+\zeta_A'(0)+\zeta_B'(0).
\eeq
 Wodzicki's formula for the multiplicative anomaly \cite{wodz87b,kass1,anom1}:
\beq \hspace*{-15mm} \delta (A,B)= \frac{\mbox{res}\left\{ \left[
\ln \sigma (A,B)\right]^2 \right\}}{2 \ \mbox{ord}\, A \
\mbox{ord}\, B \ (\mbox{ord}\, A + \mbox{ord}\, B)}, \quad \sigma
(A,B) := A^{\mbox{ord}\, B} B^{-\mbox{ord}\, A}. \eeq

At the level of Quantum Mechanics (QM), where it was originally
introduced by Feynman, the path-integral approach is just an
alternative formulation of the theory. In QFT it is much more than
this, being in many occasions {\it the} actual formulation of QFT
\cite{ramond}. In short, consider the Gaussian functional
integration \beq \int [d\Phi ] \ \exp \left\{ - \int d^D x \left[
\Phi^\dagger (x) \big( \ \ \ \ \big) \Phi (x) + \cdots \right]
\right\}  \,
  \longrightarrow  \,
\det \big( \  \ \big)^\pm,  \eeq
 and assume that the operator matrix has the following  structure
(being each $A_i$ an operator): \beq \left( \brr{cc} A_1 & A_2
\\ A_3 & A_4 \err \right) \ \longrightarrow  \,
 \left( \brr{cc} A & \mbox{} \\ \mbox{} & B \err \right), \eeq
 where the last expression is the result of diagonalizing the
operator matrix. A question now arises. What is the determinant of
the operator matrix: $ \det (AB)$ or $\det A  \cdot \det B$? This
issue has been very much on discussion \cite{1,2}.


It is difficult to give a general answer to this question, that is,
if it is possible to give a universal rule on how to choose
the right prescription, and if one can do so on mathematical
grounds only, without invoking any physical arguments. To start, we
should not forget that the issue at hand at this level is {\it regularization}.
This means, for one, that there may well be different
regularized answers that lead, after the corresponding renormalization
prescription in each case, to the same renormalized, physically meaningful
result. But the renormalization process will generically mean entering
into the physics of the problem in order to choose the right criterion.
Thus, the answer can in general only be given for the particular example considered.
There is no space here in order to enter into a more detailed discussion \cite{1,2}.

Let us just summarize by pointing out the following. First, that a number of
serious mistakes and wrong results have appeared in the literature because of
forgetting about the multiplicative anomaly. Second, that the Wodzicki
formula provides a very convenient and precise way to calculate the anomaly.
Third, that this anomaly turns often to be physically meaningful, since it
usually (but of course not always) happens that the two different regularized results obtained do
indeed lead to two different results after renormalization \cite{1,2} (therefore the
errors that have been committed in the literature, even after going through the whole process
of regularization/renormalization in a seemingly clean way). Fourth,
we know of no mathematically sound prescription in order to choose the
good regularized answer for the determinant, in general. Maybe a better answer
to this issue may be given, but it will require further investigation.


\subsection{On Determinants}
Many fundamental calculations of QFT reduce, in essence, to the
computation of the determinant of some suitable operator: at
one-loop order, any such theory reduces in fact to a theory of
determinants. The operators involved are pseudodifferential
($\Psi$DO), in loose terms `some analytic functions of differential
operators' (such as $\sqrt{1+D}$ or $\log (1+D)$, but {\it not}
$\log D$). This is explained in detail in \cite{eecmp1}. It is
surprising that this seems not to be a main subject of study among
mathematicians, in particular the determinants that involve in its
definition some kind of regularization (related to operators that
are not trace-class). This piece of calculus falls outside the scope
of the standard disciplines and even many physically oriented
mathematicians know little about this. The subject has many things
in common with divergent series but lacks any reference comparable
to the  book of Hardy \cite{hardy1}. Actually, this question  was
already addressed by Weierstrass in a way  not without problems,
since it leads to non-local contributions that cannot be given a
physical meaning in QFT. For completion, let us mention the
 well established theories of determinants for
degenerate operators, for trace-class operators in the Hilbert
space, Fredholm operators, etc. \cite{kato}

\subsection{The Chowla-Selberg Expansion Formula: Basic Aspects} From
Jacobi's identity for the $\theta-$function \beq \theta_3 (z,\tau)
:= 1+2 \sum_{n=1}^\infty q^{n^2} \cos (2nz), \qquad  q:= e^{i\pi
\tau}, \ \tau \in \mathbb{C} \eeq with \beq \theta_3 (z,\tau) =
\frac{1}{\sqrt{-i \tau}} \, e^{z^2/i\pi \tau} \, \theta_3 \left(
\frac{z}{\tau} | \frac{-1}{\tau} \right),\eeq or equivalently \beq
\sum_{n=-\infty}^\infty e^{-(n+z)^2t} = \sqrt{\frac{\pi}{t}}
\sum_{n=0}^\infty e^{-\frac{\pi^2n^2}{t}} \cos (2\pi nz), \quad z,t
\in \mathbb{C}, \ \mbox{Re}\  t > 0. \eeq In higher dimensions the
relevant expression is  Poisson's summation formula, profusely used
by Riemann in his original papers (for recent references see
\cite{gm1}, namely \beq \sum_{\vec{n} \in \mathbb{Z}^p} f(\vec{n} )
= \sum_{\vec{m} \in \mathbb{Z}^p} \tilde{f}(\vec{m} ), \eeq being
$\tilde{f}$ the  Fourier transform of $f$. An important extension of
this theory has consisted in the introduction of {\it truncated
sums} since then neither of these fundamental identities is directly
applicable \cite{elizb2}. Useful results have been obtained also in
these cases, which are very important in physical applications, in
terms of  {\it asymptotic series}.

\subsubsection{Extended CS formulas (ECS).} Consider the zeta function
(with $\mbox{Re}\  s > p/2, A > 0, \mbox{Re}\  q >0$) \beq
\hspace*{-12mm} \zeta_{A,\vec{c},q} (s) = {\sum_{\vec{n} \in
\mathbb{ Z}^p}}'
 \left[
\frac{1}{2}\left( \vec{n}+\vec{c}\right)^T A \left(
\vec{n}+\vec{c}\right)+ q\right]^{-s} = { \sum_{\vec{n} \in
\mathbb{Z}^p}}' \left[ Q\left( \vec{n}+\vec{c}\right)+ q\right]^{-s}
\eeq where the  prime indicates that the point  $ \vec{n}=\vec{0}$
is to be excluded from the sum (an inescapable condition when $
c_1=\cdots =c_p=q=0$). We can write \beq Q\left(
\vec{n}+\vec{c}\right)+ q = Q( \vec{n}) + L(\vec{n})+ \bar{q}.\eeq
\subsubsection{Case \ $q \neq 0 \ (\mbox{Re}\  q >0)$.}
Then \beq \hspace*{-14mm}\zeta_{A,\vec{c},q} (s) &=& \frac{(2\pi
)^{p/2} q^{p/2 -s}}{ \sqrt{\det A}} \, \frac{\Gamma(s-p/2)}{\Gamma
(s)} +
 \frac{2^{s/2+p/4+2}\pi^s q^{-s/2 +p/4}}{\sqrt{\det A} \
\Gamma (s)}  \nn\\ && \hspace*{-14mm} \times {\sum_{\vec{m} \in
\mathbb{Z}^p_{1/2}}}\hspace*{-2mm}' \cos (2\pi
 \vec{m}\cdot \vec{c}) \left( \vec{m}^T A^{-1} \vec{m}
\right)^{s/2-p/4} \, K_{p/2-s} \left( 2\pi \sqrt{2q \,
 \vec{m}^T A^{-1} \vec{m}}\right),
\eeq an original expression that we have labeled as [ECS1]. After
detailed inspection, it is easy to see here that the pole at $
s=p/2$, and its corresponding residue \beq \mbox{Res}_{s=p/2} \,
\zeta_{A,\vec{c},q} (s) = \frac{(2\pi)^{p/2}}{\Gamma(p/2)}\, (\det
A)^{-1/2}, \eeq are explicitly given in the formula, which has in
all the following properties. \ben
 \item It yields the (analytical continuation of) the multidimensional
zeta function in terms of an {\it exponentially convergent}
multiseries, valid in the {\it whole} complex plane
\item It exhibits singularities ({\it simple poles}) of the meromorphic
continuation---with the corresponding {\it residua}---{\it
explicitly}.
\item The only condition on the matrix, $A$, is that it must correspond
to a {\it (non negative) quadratic form}, $Q$. The vector $\vec{c}$
is {\it arbitrary}, while $q$ is  (to start) any non-negative
constant.
\item $K_\nu$ is the modified Bessel function of the second kind
and the subindex in $\mathbb{Z}^p_{1/2}$ means that only {\it half}
of the vectors $\vec{m} \in \mathbb{Z}^p$ participate in the sum.
E.g., if we take an index $ \vec{m} \in \mathbb{Z}^p$ we must then
exclude $ -\vec{m}$, a simple criterion being:  one may select those
vectors in $\mathbb{Z}^p \backslash \{ \vec{0} \}$ whose {\it first
non-zero component is positive}. \een

\subsubsection{Case \ $ c_1=\cdots =c_p=q=0$.}
 This case is a true extension of CS; we will here consider the diagonal subcase
 only \cite{eejpa2}
\beq && \hspace*{-22mm} \zeta_{A_p} (s)  =  \frac{2^{1+s}}{\Gamma
(s)} \, \sum_{j=0}^{p-1} \left(\det{A_j} \right)^{-1/2} \left[
\pi^{j/2} a_{p-j}^{j/2-s} \right.  \Gamma \left( s-\frac{j}{2}
\right) \,
\zeta_R(2s-j) \nn \\
 && \hspace*{-19mm}  +4 \pi^s a_{p-j}^{\frac{j}{4}-\frac{s}{2}}
\sum_{n=1}^\infty {\sum_{\vec{m}_j \in \mathbb{Z}^j}}\hspace*{-2mm}
' n^{j/2-s} \left(\vec{m}_j^t
A_j^{-1}\vec{m}_j\right)^{s/2-j/4}\left. K_{j/2-s}\left(2\pi n
\sqrt{ a_{p-j} \vec{m}_j^t A_j^{-1}\vec{m}_j}\right) \right], \eeq
an expression that truly extends the CS formula and we have labeled
as [ECS3d] \cite{eejpa2}.

\section{On Zeta Function Regularization}
\subsection{Some Considerations on Zeta Regularization}
Regularization and renormalization procedures are essential issues
in contemporary physics \cite{renorm1}.  Among the different
methods, zeta function regularization---obtained by analytic
continuation  in the complex plane of the zeta function of the
relevant physical operator in each case---is one of the most
beautiful of all. Use of this procedure yields the vacuum energy
corresponding to a quantum physical system, with constraints of very
different nature. The case of moving boundaries seems to present
quite severe difficulties, though some promising approach to deal
with them has appeared \cite{haee1}. Let the Hamiltonian operator,
$H$, of our quantum system to have a spectral decomposition of the
form (think as simplest case in a quantum harmonic oscillator): $\{
\lambda_i, \varphi_i \}_{ i\in I}$, with $I$ some set of indices (it
can be discrete, continuous, mixed, or multiple). The quantum vacuum
energy is obtained as follows \cite{zb12k}
\begin{eqnarray}  \hspace*{-8mm} E/\mu = \sum_{i\in I}\langle \varphi_i, (H/\mu)
\varphi_i \rangle = \mbox{Tr}_{\zeta} H/\mu  = \left.  \sum_{i\in I}
(\lambda_i/\mu)^{-s} \right|_{s=-1} \hspace*{-3mm}= \zeta_{H/\mu}
(-1),
\end{eqnarray}
 where $\zeta_A$ is the zeta function corresponding to the operator $A$,
and the equalities are in the sense of analytic continuation (since,
generically, the Hamiltonian operator will not be of the trace
class). Actually, this $\zeta-$trace is {\it no trace in the usual
sense}. It is highly non-linear, as often explained by the author
\cite{eejhep1}. Some colleagues are however unaware of this fact,
which has lead to very serious mistakes and erroneous conclusions in
the literature.

The formal sum over the eigenvalues is usually ill defined and the
last step involves analytic continuation, inherent with the
definition of the zeta function itself. Also, an unavoidable
renormalization parameter, $\mu$, with the dimensions of mass,
appears in the process, in order to render the eigenvalues of the
resulting operator dimensionless, so that the corresponding zeta
function can actually be defined. For lack of space, we shall not
discuss those basic details here, which are at the starting point of
the whole renormalization procedure. The mathematically
simple-looking relations above involve deep physical concepts, no
wonder that understanding them has taken several decades in the
recent history of quantum field theory.

\subsection{On the Zero Point Energy and the Casimir Force} In an
ordinary QFT, one cannot give a meaning to the {\it absolute} value
of the zero-point energy, and any physically measurable effect comes
as an energy {\it difference} between two situations, such as a
quantum field satisfying BCs on some surface as compared with the
same in its absence, or one in curved space as compared with the
same field in flat space, etc. This difference is the Casimir
energy:
$E_C = E_0^{BC} - E_0 =  \frac{1}{2} \left( \mbox{tr } H^{BC} -
\mbox{tr } H \right)$.
But here a problem appears. Imposing mathematical boundary
conditions (BCs) on physical quantum fields turns out to be a highly
non-trivial issue. This was discussed in detail in a paper by
Deutsch and Candelas \cite{dc79}. These authors quantized em and
scalar fields in the region near an arbitrary smooth boundary, and
calculated the renormalized vacuum expectation value of the
stress-energy tensor, to find out that the energy density diverges
as the boundary is approached. Therefore, regularization and
renormalization did not seem to cure the problem with infinities in
this case and an infinite {\it physical} energy was obtained if the
mathematical BCs were to be fulfilled. However, the authors argued
that  surfaces have non-zero depth, and its value could be taken as
a handy dimensional cutoff in order to regularize the infinities.
Just two years after Deutsch and Candelas' work, Kurt Symanzik
carried out a rigorous analysis of QFT
 in the presence of boundaries \cite{ks81}. Prescribing the value
of the quantum field on a boundary means using the Schr\"odinger
representation, and Symanzik was able to show rigorously that such
representation exists to all orders in the perturbative expansion.
He showed also that the field operator being diagonalized in a
smooth hypersurface differs from the usual renormalized one by a
factor that diverges logarithmically when the distance to the
hypersurface goes to zero. This requires a precise limiting
procedure and  point splitting to be applied. In any case, the issue
was proven by him to be perfectly meaningful within the domains of
renormalized QFT. In this case the BCs and the hypersurfaces
themselves were treated at a pure mathematical level (zero depth) by
using Dirac delta functions.

Not long ago, a new approach to the problem has been postulated
\cite{bj1}. BCs on a field, $\phi$, are enforced on a surface, $S$,
by introducing a scalar potential, $\sigma$, of Gaussian shape
living on and near the surface. When the Gaussian becomes a delta
function, the BCs (Dirichlet here) are enforced: the delta-shaped
potential kills {\it all} the modes of $\phi$ at the surface. For
the rest, the quantum system undergoes a full-fledged QFT
renormalization, as in the case of Symanzik's approach. The results
obtained confirm those of \cite{dc79} in the several models studied
albeit they do not seem to agree with those of \cite{ks81}. They
seem to be also in contradiction with the ones quoted in the usual
textbooks and review articles dealing with the Casimir effect
\cite{cb1}, where no infinite energy density when approaching the
Casimir plates has been reported. This has been extended by the
author using methods of Hadamard regularization, what seems to be a
new important development in this direction \cite{eehad1}.

\section{Quantum Vacuum Fluctuations, Zeta Regularization,
and the Cosmological Constant}
\subsection{Vacuum Energy Fluctuations and the Cosmological Constant}
The issue of the cc has got renewed thrust from the observational
evidence of an acceleration in the expansion of our Universe,
initially reported by two different groups \cite{perlries}. There
was some controversy on the reliability of the results obtained from
those observations and on its precise interpretation, but after new
data was gathered, there is now consensus among the community of
cosmologists that, in fact, an acceleration is there, and that it
has the order of magnitude obtained in the above mentioned
observations \cite{ries2,Carroll1,Carroll2}. As a consequence, many
theoreticians have urged to try to explain this fact, and also to
try to reproduce the precise value of the cc coming from these
observations \cite{sts1,shs1,mon1}.

As crudely stated by Weinberg \cite{wei2}, it is more difficult to
explain why the cc is so small but non-zero, than to build
theoretical models where it exactly vanishes \cite{cwp1}. Rigorous
calculations performed in quantum field theory on the vacuum energy
density, $\rho_V$, corresponding to quantum fluctuations of the
fields we observe in nature, lead to values that are many orders of
magnitude in excess of those allowed by observations of the
space-time around us. Energy always gravitates \cite{eqpri1},
therefore the energy density of the vacuum,  more precisely, the
vacuum expectation value of the stress-energy tensor $ \langle
T_{\mu \nu} \rangle \equiv - {\cal E} g_{\mu \nu} $ appears on the
rhs of Einstein's equations: \beq
R_{\mu\nu}-\frac{1}{2}g_{\mu\nu}R=-8\pi G(\tilde{T}_{\mu\nu}-{\cal
E}g_{\mu\nu}). \eeq It affects cosmology: $\tilde T_{\mu \nu}$
contains excitations above the vacuum, and is equivalent to a {\it
cc} $ \Lambda =8\pi G{\cal E}$.
 Recent observations yield \cite{sdsscol1}
 \beq \Lambda_{\mbox{\small obs}}  \ = \ (2.14\pm 0.13 \times
10^{-3}\ \hbox{eV})^{4} \quad \sim \ 4.32 \times 10^{-9}\
\hbox{erg/cm}^3 \nonumber \eeq It is an old idea that the cc gets
contributions from zero point fluctuations  \cite{zeld1} \beq  E_0 \
= \ \frac{\hbar\, c}{2} \sum_n \omega_n, \qquad \omega = k^2 +
m^2/\hbar^2, \ \ k =2 \pi /\Lambda. \eeq  Evaluating in a box and
putting a cut-off at maximum $k_{max}$ corresponding to reliable QFT
physics (e.g., the Planck energy) \beq   \rho  \ \sim \ \frac{\hbar
\, k_{\mbox{\footnotesize Planck}}^4}{16 \pi^2} \ \sim \ 10^{123}
\rho_{\mbox{\small obs}}.\eeq

 Assuming one will be able to prove (in the future) that the ground
value of the cc is {\it zero} (as many suspected until recently), we
will be left with this {\it incremental value} coming from the
topology or BCs. This sort of two-step approach to the cc is
becoming more and more popular recently as a way to try to solve
this very difficult issue \cite{ccr2}. We have then to see, using
different examples, if this value acquires the  correct order of
magnitude ---corresponding to the one  coming from the observed
acceleration in the expansion of our universe--- under some
reasonable conditions. We pursue a quite simple and primitive idea,
related with the {\it global} topology of the universe \cite{ct1}
and in connection with the possibility that a faint scalar field
pervading the universe could exist. Fields of this kind are
ubiquitous in inflationary models, quintessence theories, and the
like. In other words, we do not pretend  to solve the old problem of
the cc, not even to contribute significantly to its understanding,
but just to present simple and usual models which show that the
right order of magnitude of (some contributions to) $\rho_V$ which
lie in the precise range deduced from the astrophysical observations
are not difficult to get. In different words, we only address here
the 'second stage' of what has been termed by Weinberg \cite{wei2}
the {\it new} cc problem.

\subsection{Vacuum Energy Contribution in Different Models}
\subsubsection{Simple model with large and small compactified
dimensions.} We assume the existence of a scalar field  extending
through the universe and calculate the contribution to the cc from
the Casimir energy density of this field, for some typical boundary
conditions. Ultraviolet contributions will be set to zero by
some mechanism of a fundamental theory. We assume
the existence of both large and small dimensions (the total number
of large spatial coordinates being always three), some of which  may
be compactified, so that the global topology of the universe may
play an important role \cite{ct1,sokol1,css-mfo,elif12,banks1}.
We know \cite{zb12k}  that the range
of orders of magnitude of the vacuum energy density for
common possibilities is not widespread (may only differ by
a couple of digits) and one can deal with two simple situations:
a scalar field with periodic BCs or spherically
compactified \cite{eenc,eeqfext05}). The contribution of the vacuum energy of a
small-mass scalar field, conformally coupled to gravity, and coming
from the compactification of some small (2 or 3) and some large (1
or 2) dimensions ---with compactification radii of the order of 10
to 1000 the Planck length in the first case and of the order of the
present radius of the universe, in the second--- lead to values that
compare well with observational data, in order of magnitude, but with
the wrong {\it sign}.

\subsubsection{Braneworld models.}
An important issue in all the previous analysis is the specific {\it
sign} of the resulting force. For scalar fields and the usual
compactifications or BCs it is impossible to get the right sign
corresponding to the accelerated expansion of the universe. However,
in braneworld models and others involving supergravitons and fermion
fields we have been able to prove that the appropriate sign can be
obtained under quite natural conditions.

Braneworld theories may hopefully
solve both the hierarchy problem and the cc problem. The bulk
Casimir effect can play an important role in the construction
(radion stabilization) of braneworlds. We have calculated the bulk
Casimir effect (effective potential)  for conformal and for  massive
scalar fields \cite{enoo2003}. The bulk is a 5-dim AdS or dS space,
with 2 (or 1) 4-dim dS branes (our universe). The results obtained
are quite consistent with observational data. A difficulty in this case,
however, is the comparison of the vacuum energy density obtained in
five dimension with the one corresponding to four dimensions. Even more,
six dimensional models are very on fashion now and problems of this kind
pop up there too \cite{emn}.

\subsubsection{Supergraviton theories} We have also computed
the effective potential for some multi-graviton models with
supersymmetry \cite{ss1}. In one case, the bulk is a flat manifold
with the torus topology $\mathbb{R} \times \mathbb{T}^3$, and it can
be shown that the induced cc can  be rendered {\it positive} due to
topological contributions \cite{cezplb05}. Previously, the case of
$\mathbb{R}^4$ had been considered. In the multi-graviton model the
induced cc can indeed be positive, but only if the number of massive
gravitons is sufficiently large, what is not easy to fit in a
natural way. In the supersymmetric case, however, the cc turns out
to be positive just by imposing anti-periodic BC in the fermionic
sector. An essential issue in our model is to allow for
non-nearest-neighbor couplings.

For the torus topology we have got the topological contributions to
the effective potential to have always a fixed sign, which depends
on the BC one imposes.  They are negative for periodic fields, and
positive for anti-periodic ones. But topology provides then a
mechanism which, in a natural way, permits to have a positive cc in
the multi-supergravity model with anti-periodic fermions. The value
of the cc is regulated by the corresponding size of the torus. We
can most naturally use the minimum number, $N = 3$, of copies of
bosons and fermions, and show that ---as in the first, much more
simple example, but now with the right sign!--- within our model the
observational values for the cc can be approximately matched, by
making quite reasonable adjustments of the parameters involved. As a
byproduct, the results that we have obtained \cite{cezplb05} might
also be relevant in the study of electroweak symmetry breaking in
models with similar type of couplings, for the deconstruction issue.
\medskip

\noindent{\bf Acknowledgments.} Based on work done in part in
collaboration with G. Cognola, S. Nojiri, S.D. Odintsov and S.
Zerbini. The financial support of DGICYT (Spain), project
FIS2006-02842, and of AGAUR (Generalitat de Catalunya), project
2007BE-1003 and contract 2005SGR-00790, is gratefully acknowledged.

\end{document}